\documentclass[11pt,a4paper]{article}
\pdfoutput=1
\usepackage{a4wide}
\usepackage{jheppub}

\usepackage{epsfig}
\usepackage{latexsym}
\usepackage{amsfonts}
\usepackage{amsmath}
\usepackage{amsthm}
\usepackage{amssymb}
\usepackage{amsbsy}
\usepackage{multirow}
\usepackage{slashed}
\usepackage{color}
\usepackage{mathrsfs}
\usepackage{subfigure}
\usepackage[font={footnotesize,sl},bf]{caption}
\usepackage{wrapfig}

\newcommand{\e}{\textrm{e}}

\newcommand{\w}{\wedge}

\newsavebox{\uuunit}
\sbox{\uuunit}
    {\setlength{\unitlength}{0.825em}
     \begin{picture}(0.6,0.7)
        \thinlines
        \put(0,0){\line(1,0){0.5}}
        \put(0.15,0){\line(0,1){0.7}}
        \put(0.35,0){\line(0,1){0.8}}
       \multiput(0.3,0.8)(-0.04,-0.02){12}{\rule{0.5pt}{0.5pt}}
     \end {picture}}

\newcommand{\beq}{\begin{eqnarray}}
\newcommand{\eeq}{\end{eqnarray}}

\newcommand{\be}{\begin{equation}}
\newcommand{\ee}{\end{equation}}
\newcommand{\bea}{\begin{eqnarray}}
\newcommand{\eea}{\end{eqnarray}}
\newcommand{\bean}{\begin{eqnarray*}}
\newcommand{\eean}{\end{eqnarray*}}

\def\to{\rightarrow}

\def\sF{{{ F}\!\!\!\!\hskip.8pt\hbox{\raise1pt\hbox{/}}\,}}
\def\som{{{ \omega}\!\!\!\!\hskip.8pt\hbox{\raise1pt\hbox{/}}\,}}
\def\sJ{{{\rm J}\!\!\!\!\hskip.8pt\hbox{\raise1pt\hbox{/}}\,}}

\newcommand{\bdm}{\begin{displaymath}}
\newcommand{\edm}{\end{displaymath}}

\begin{document}

\title{Observations on fluxes near anti-branes}
\author{Diego Cohen-Maldonado$^{a}$,}
\author{Juan Diaz$^b$,}
\author{Thomas Van Riet$^{b}$,}
\author{Bert Vercnocke$^{a}$}
\affiliation{  $^{a}$ Institute of Physics, University of Amsterdam, Science Park,\\ 
	Postbus 94485, 1090 GL Amsterdam, The Netherlands }
\affiliation{ $^{b}$ Instituut voor Theoretische Fysica, K.U. Leuven,\\
	Celestijnenlaan 200D B-3001 Leuven, Belgium }

\emailAdd{d.b.cohenmaldonado@uva.nl}
\emailAdd{juan@itf.fys.kuleuven.be}
\emailAdd{thomasvr@itf.fys.kuleuven.be}
\emailAdd{bert.vercnocke@uva.nl}

\abstract{We revisit necessary conditions for gluing local (anti-)D3 throats into flux throats with opposite charge.  These consistency conditions typically reveal singularities in the 3-form fluxes whose meaning is being debated. In this note we prove, under well-motivated assumptions, that unphysical singularities can potentially be avoided when the anti-branes polarise into spherical NS5 branes, with a specific radius. If a consistent solution can then indeed be found, our analysis seems to suggests a rather large correction to the radius of the polarization sphere compared to the probe result. We furthermore comment on the gluing conditions at finite temperature and point out that one specific assumption of a recent no-go theorem can be broken if anti-branes are indeed to polarise into spherical NS5 branes at zero temperature. 
}

\maketitle

\section{Introduction}
	Supersymmetric throat geometries supported by fluxes are stable string theory solutions with important applications for holography, flux compactifications and black holes. One explicit method to break the supersymmetry in the throat while preserving classical stability is adding branes to the flux background that carry charges opposite to the charge dissolved in flux \cite{Maldacena:2001pb, Kachru:2002gs, Kachru:2003aw, Argurio:2007qk, Bena:2012zi}, from here on referred to as ``anti-branes". We focus on anti-D3 branes in the Klebanov--Strassler (KS) throat as first studied in \cite{Kachru:2002gs}. If the anti-brane charge $p$ is small compared to the background RR-flux $M$, then their addition can be seen as a small perturbation of the KS throat. At the same time, the limit of small anti-brane charge is what guarantees metastability, at least at probe level \cite{Kachru:2002gs}. The decay channel is the annihilation of the $p$ anti-D3 branes with some of the background NSNS flux. If the corrections to the probe result come with positive powers of $p/M$ it guarantees the self-consistency of the approach as long as $p/M\ll1$. 
	
	If the above reasoning is to hold, then the results of \cite{Kachru:2002gs} suggest that the supergravity solution describing the metastable state should be spherical NS5 branes wrapping a contractible two-cycle of finite size. These spherical NS5 branes carry no monopole NS5 brane charge but the original $p$ anti-D3 charges instead. This supergravity picture should hold in the following limit:	
	\be
	g_s \ll1\,,\qquad g_sp \gg1\,,\qquad g_sM \gg1\,,
	\ee
	and at the same time $p/M\ll1$ as required for stability. The above limit ensures that string loop corrections can be ignored and that the length scales of the tip of the KS throat and of the contractible 2-cycle are large in string units such that higher  derivate corrections are also suppressed.  	
	Recently interesting arguments have been presented that for $p=1$ the metastability is not threatened by backreaction \cite{Michel:2014lva, Bergshoeff:2015jxa}. A single anti-brane is however beyond the scope of this paper since it cannot be understood within the 10D supergravity limit. The complementary regime of $p/M$ of order unity or larger cannot be regarded as a perturbation to the KS throat and the would-be supergravity solution in this case can better be thought of as being AdS$_5\times$S$^5$ perturbed by $M$ units of three-form flux \cite{Polchinski:2000uf}. We do not go into that limit in this work since it does not correspond to meta-stable supersymmetry breaking by perturbatively small amounts as originally intended in \cite{Kachru:2002gs}.

The first attempts for constructing the supergravity solutions (in certain approximate schemes)  revealed singularities which were claimed to be different from the usual singularities associated to brane sources \cite{Bena:2009xk, McGuirk:2009xx}. The singularity is such that the scalar $e^{-\phi}|H_3|^2$, 
which  gives the  contribution of the $H_3$ flux to the energy-momentum tensor in Einstein frame,  diverges near the sources. Although it is tempting to interpret this divergence as the self-energy of the NS5 brane, it was claimed this is not the correct interpretation, because the orientation of the fluxes and the magnitude of the divergence was not right for NS5 branes. One might worry that this mismatch is inherent to the approximations of the original papers \cite{Bena:2009xk, McGuirk:2009xx}, but in fact the singularity was shown to persist beyond those approximations \cite{Gautason:2013zw, Blaback:2014tfa}. Nonetheless, that interpretation of the singularity as unphysical is incomplete since the computations of \cite{Gautason:2013zw, Blaback:2014tfa} assume genuine anti-D3 branes instead of spherical NS5 branes carrying anti-D3 charges. One could therefore speculate that when one looks instead for supergravity solutions describing spherical NS5 branes at the tip of the throat, one finds acceptable singularities. This is the first point we investigate in this paper and we find that, under well motivated assumptions, certain fields can be tuned near the horizon such that the singularity corresponds to the usual divergent self-energy of the NS5 brane.  There is no guarantee that this tuning is possible in the sense that the UV can be glued consistently to the IR, but at least we find that all earlier no-go theorems against that gluing can be circumvented.

Second we investigate the configuration at finite temperature \cite{Gubser:2000nd}.  The temperature acts as an IR cut-off in field theory. If there is a mechanism in string theory that can resolve the singularity or turn it into a physical divergence (such as with brane polarisation), one expects the singularity to be shielded at finite temperature.  
Numerical evidence has shown that this hope was in vain \cite{Bena:2012ek, Buchel:2013dla} at least for anti-D3 branes that are smeared over the finite tip at the bottom of the throat or for localised anti-D6 branes \cite{Bena:2013hr}. In addition, a no-go theorem was found for localised anti-D$p$ branes with $p\leq 6$ \cite{Blaback:2014tfa}: the would-be  supergravity solution will develop a divergent flux density $e^{-\phi} |H_3|^2$ at the horizon.  However, as all no-go results, the theorem is only as strong as its assumptions.  In this paper we argue that NS5 polarization is a priori consistent with relaxing one assumption in \cite{Blaback:2014tfa} on the boundary conditions. If finite $T$ solutions exist, our results should then be a useful lead on how to pick boundary conditions near the brane sources. In fact, some progress on the existence of smooth finite $T$ solutions was reported in \cite{Hartnett:2015oda} and we verify that indeed this boundary condition was used. Still the construction of \cite{Hartnett:2015oda} misses a compact `A'-cycle, which is crucial for the physics of anti-brane metastable states. Without such a compact A-cycle the construction of smooth finite temperature solutions was also argued earlier in \cite{Freedman:2000xb}. 

We start the remainder of this note with a review of the main results of \cite{Blaback:2014tfa}. 
	Then we analyse the $H_3$ flux density. At zero temperature, we review the singular flux for anti-D3 brane boundary conditions in the IR region of the flux throat and extend the analysis to spherical NS5-branes carrying anti-D3 charge. We show the flux energy density is again singular, but it is possible to obtain the appropriate power of divergence expected for a local NS5-source. This turns out to be only possible at a NS5 radius which scales as $R\sim \sqrt{p/M}$ for small $p/M$, which differs from the probe prediction $R\sim p/M$. We then heat up the background, and elaborate on a caveat in the arguments of \cite{Blaback:2014tfa} and discuss under which conditions the singularity can be cloaked by a smooth horizon. Finally, we discuss and interpret our results.

\section{Gluing IR to UV at \texorpdfstring{$T=0$}{}} \label{gluing}
	We recall the formalism of \cite{Blaback:2014tfa}. This formalism extends the results in \cite{Gautason:2013zw} and relates the boundary conditions near the anti-brane in the IR to the generalized ADM mass, which is measured in the UV.  These consistency relations are necessary (but not sufficient) for the existence of well-behaved interpolating solutions. They have proven useful to demonstrate the absence of solutions in many non-trivial circumstances, which are otherwise only amenable to heavy numerical techniques.	
	
	We focus the discussion on throat geometries supported by 3-form fluxes in which anti-D3 branes are added at the tip. The example to have in mind is the Klebanov--Strassler throat \cite{Klebanov:2000hb}.  
	The would-be solution takes the following form in 10D Einstein frame:
	\begin{align}
	&\mathrm{d} s^2_{10} = \e^{2A}\Bigl(-e^{2f} \mathrm{d}t^2 + \delta_{ij} \mathrm{d}x^i \mathrm{d}x^j\Bigr) + \mathrm{d} s^2_{6}\,,\nonumber\\
	& C_{4}= \tilde \star_{4} \alpha ,\nonumber\\
	& H_3 = -\e^{\phi-4A-f} \star_6\left((\alpha+\alpha_0) F_3 +  X_{3}\right)\,. \label{H3ansatz} 
	\end{align}
	The functions $\phi, A, f, \alpha$ only depend on the internal coordinates and $\alpha_0$ is a constant.
	The horizon is located at $e^{2f} =0$. At this point the throat metric $\mathrm{d} s_6^2$ is completely general. 
	We use notation with tildes for metrics and Hodge duals without any warp factor $e^{2A}$ nor $e^{2f}$. For instance $\tilde g_{\mu\nu} = \eta_{\mu\nu}$ is the Minkowski metric.
	
	We require the throat geometries to have a $3$-dimensional subspace non-trivial in homology, which we  call the A-cycle, and $F_3$ to have a fixed quantised flux $\int_A F_3 = 4\pi^2 M$.
	This piece of information is crucial to describe the backreaction of  anti-branes in the Klebanov-Strassler background. The A-cycle is present in the UV limit of the supergravity solution \cite{Bena:2009xk} and was essential in proving the presence of the  original non-perturbative brane-flux instabilities \cite{Kachru:2002gs} and the singularity \cite{Blaback:2014tfa}. For a relatively low number of anti-branes  $p$ such that  $p/M \ll 1$ we expect the topology to remain the same and the A-cycle to be present.
	
	The $3$-form fluxes $H_3$ and $F_3$ are closed and take values inside the $6$-dimensional throat geometry. 
	Then $H_3$  is the most general form that solves the $B_2$ equation of motion
	\begin{equation}\label{B2EOM}
	\mathrm{d} (\e^{-\phi}\star_{10} H_3) = -F_{5}\wedge  F_{3}\, ,
	\end{equation}
	provided that $X_3$ is closed as well.	The ansatz for the $H_3$ flux \eqref{H3ansatz} seems to have a redundancy, as any shift in $X_3$ along $F_3$
        shifts the constant $\alpha_0$. We fix this redundancy by demanding that $\int _A X_{3}=0$. 
	We will furthermore fix the gauge for $C_4$ (and hence $\alpha$) such that $\alpha_0 = 0$. This is the gauge used in \cite{Bena:2009xk, Dymarsky:2011pm}, whose results we use below.
	
	The key observation of \cite{Blaback:2014tfa}, which we repeat in the appendix, is that the following $9$-form
	\be \label{9form}
	\mathcal{B} = -C_4\w F_5 -\tilde{\star_4}1\w B_2\w X_3 +\star_{10} \mathrm{d}\Bigl(\phi - 4A -f \Bigr)\,
	\ee
	integrates over a spacetime volume $\mathcal M$ to
	\be\label{Onshellbraneaction}
        \frac 1 {\tilde{v}_4}\oint_{\partial {\cal M}} \mathcal{B} = M_{ADM}\,,
	\ee
	if the closed surface $\partial\mathcal M$ encapsulates the anti-brane sources; otherwise the integral vanishes. In this expression, $\tilde{v}_4$ is the four-dimensional volume. This is very much like Gauss' law in electrodynamics with the role of the electric charge played by the generalized ADM mass. When supersymmetry is broken by $p$ anti-branes,  $M_{ADM}$ is proportional to $p$ and positive \cite{Blaback:2014tfa}. This condition can then be used to constrain the behavior of the supergravity fields near the source by letting the closed surface approach the anti-brane horizon. We want to stress that this integral has only been shown to reproduce the  ADM mass for the perturbative solution of \cite{Bena:2009xk}. We come back to more general asymptotics allowing  $p/M \gg 1$ in section \ref{ssec:Smarr}. 
	
\begin{figure}[ht!]
		\begin{center}
		 \includegraphics[width=.55\textwidth]{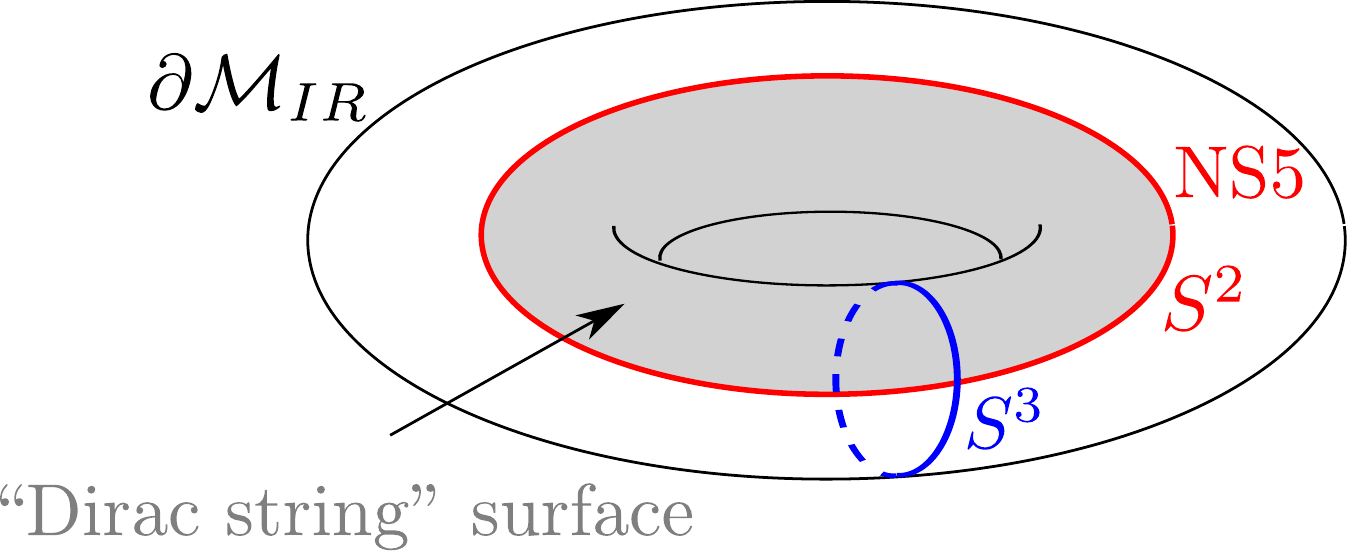}
		 \hspace{.1\textwidth}
		 \includegraphics[width=.33\textwidth]{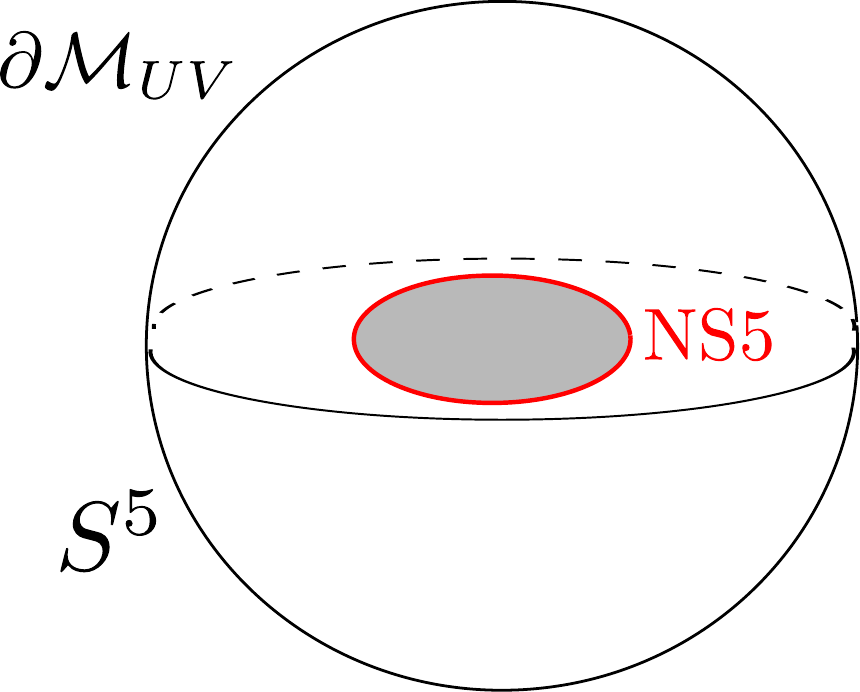}
		 \caption{Boundaries $\partial {\cal M}$, with 4d Minkowski directions suppressed. Left: the IR surface that encapsulates the NS5 corresponds to an $S^3 \times S^2$ close to the brane. The spherical NS5 is drawn symbolically as a red circle. We can choose  a patch such that the $B_2$ gauge field  is well-defined everywhere except in they gray surface stretching  from one side of the NS5-brane to another.  Right: the UV boundary can be taken to be $S^5$.}
		 \label{fig:Torus}
		 \end{center}	
		\end{figure}

	The above formalism needs to be altered in order to apply it to spherical NS5 branes. The reason is that we intend to integrate the 9-form (\ref{9form}) on a surface $\partial{\cal M}_{IR}$ infinitesimally close to the spherical NS5 horizon. But along this surface $B_2$ is not everywhere defined since an NS5 acts as a magnetic source for $B_2$. This would imply that we need to compute the contribution from a surface of Dirac ``strings'' stretching from one side of the spherical NS5 to the opposite side, as depicted in Figure \ref{fig:Torus}.\footnote{In a first version of this paper we missed that contribution and we are grateful to J. Polchinski for pointing that out. See also comments in \cite{Polchinski:2015bea}.}
	
	 Luckily this complication can be avoided by using $B_6$ as the fundamental potential instead of $B_2$. Going through an analogous computation as in \cite{Blaback:2014tfa}, but using $B_6$, one finds:
	\be \label{9form2}
	\mathcal{B} = -C_4\w F_5 - B_6 \w H_3 +\star_{10} \mathrm{d}\Bigl(\phi - 4A -f \Bigr)\,,
	\ee
	where we made use of
	\be\label{B6def}
	\e^{-\phi}\star H_3 \equiv \mathrm{d} B_6 -C_4\w F_3\,,
	\ee
	such that
	\be\label{dB6}
	\mathrm{d} B_6 = -\tilde{\star_4}1 \w X_3\,.
	\ee
	Note that we ignore a possible closed  but not exact piece in $X_3$. Such a harmonic piece would give a non-zero bulk contribution $\int_{\cal M} H_3 \wedge \tilde \star_4 1 \wedge X_3$ to the ADM mass (see Appendix \ref{app:boundary} for more information). This harmonic contribution to the ADM mass would be similar to the mechanism to smooth support black hole  microstate geometries \cite{Gibbons:2013tqa}. We will explore this connection in more detail in  a forthcoming publication.

	\section{Flux divergences}\label{divergences}

	We focus on the density of the NS-NS 3-form flux as it appears in the energy-momentum tensor. From the ansatz \eqref{H3ansatz} we find
	\be\label{Hdensity}
	\e^{-\phi}|H_3|^2 =  e^{\phi - 8A -2f}|\alpha F_3 + X_3|^2\,.
	\ee
	It is the aim of this paper to infer whether this quantity stays finite near the source or not, by using consistency relations for gluing the IR solution to the UV solution.

	\subsection{Anti-D3 boundary condition}
Consider eq.~(\ref{Hdensity}) and take zero temperature ($e^{2f}=1$). Close to the anti-D3 brane, we can use the near-D3 solution of the appendix, eq.\ \eqref{eq:pbranemetric}. The factor $e^{-8A}$ is expected to blow up whereas $\phi$ remains finite. The metric transverse to the brane scales as $\mathrm{d}s^2_6 = e^{-2A} \widetilde{\mathrm{d}s}^2_6$, such that
	\be\label{Hdensity2}
	\e^{-\phi}|H_3|^2 \sim e^{-2A}{|\alpha F_3 + X_3|}_{\widetilde {\mathrm{d}s}_6^2}^2\,,
	\ee
	where the subscript indicates  we take the contraction without warp factors using $\widetilde{\mathrm{d}s}^2_6$. Hence unless the combination $\alpha F_3+X_3$ vanishes at the position of the brane, we find a singular $H_3$ density scaling. This is the infamous 3-form singularity encountered in many places in the literature. 
	
	One can argue that the combination $\alpha F_3 + X_3$ indeed does not vanish at the anti-D3-brane position using eq.\ (\ref{Onshellbraneaction}). As we are dealing with anti-D3 branes, we are free to use the (\ref{9form}) boundary term, since we do not expect a contribution from Dirac strings. For the standard (anti-)D3 boundary condition, the last term (whole bracket) in  (\ref{9form}) does not contribute. Hence $X_3$ and $\alpha$ cannot both vanish since $M_{ADM}>0$. As $F_3$ has a non-vanishing component along the A-cycle, the combination  $\alpha F_3 + X_3$ is nonzero by construction and hence introduces a three-form singularity. Note that for a D3-brane boundary condition with $M_{ADM}=0$, this issue is not present.
	
	A possible way out would be that $F_3$ and $X_3$ vanish at the position of the anti-D3 brane, while $\alpha$ is  non-zero (see the discussion in \cite{Gautason:2013zw}). This is a priori not impossible since only the integral along the A-cycle of $F_3$ equals $4\pi^2M$ and the flux could be distributed inhomogeneously along that cycle. If one tracks down what is needed for a regular $H$-flux density, one finds that the charge density in the fluxes has to scale to zero near the source as 
	\be\label{chargedensity1}
	H_3\wedge F_3 \sim e^{4A}\star_6 1\,.
	\ee
	We consider this to be an unphysical boundary condition, since the anti-brane attracts both gravitationally and electromagnetically the charges dissolved in flux; therefore one expects that the maximal value for $H_3\wedge F_3$ is reached near the source, instead of going to zero. We leave it for further research to find a full mathematical proof of this.

	\subsection{NS5 boundary condition}
	In the probe limit, the anti-D3 brane is unstable towards puffing up into an NS5-brane that wraps a contractible 2-cycle inside the A-cycle \cite{Kachru:2002gs}. This NS5 brane carries no NS5 monopole charge, but rather anti-D3 charge through flux on its worldvolume.  Since the probe NS5-branes can sit at a classically stable position, one expects a consistent supergravity solution with an NS5 brane boundary condition in the IR. We argue now that this possibility can indeed not be excluded from the gluing conditions.
	
	With NS5 boundary conditions at the brane position, the four-dimensional transverse metric scales as $e^{-6A}$, while $e^{\phi}$ scales as $e^{-4A}$. Three non-trivial conditions have to be met in order for the $H_3$ density  (\ref{Hdensity}) to be consistent with an NS5 brane (the interested reader can corroborate them using the expressions in Appendix \ref{app:sugra}):
	\begin{enumerate}
		\item Near the NS5 source, the $H_3$ density should be singular of a specific degree: $e^{-\phi} |H_3|^2 \sim \e^{-2A}$.
		\item Near the source, the dominating legs of $H_3$ are perpendicular to the NS5 worldvolume.
		\item $\oint_{\partial {\cal M}}\mathcal{B}$ has to be finite and positive by (\ref{Onshellbraneaction}),  with ${\cal M}$ a spacetime volume encapsulating the source.
	\end{enumerate}
	The ansatz \eqref{H3ansatz} and the second requirement imply that $\alpha F_3 + X_3$ must have two legs on the NS5 worldvolume and one transverse leg. Together with the local metric scaling of an NS5 brane  this implies
	\be\label{Hdensity3}
	\e^{-\phi}|H_3|^2 =  \e^{-10 A}|\alpha F_3 + X_3|_{\widetilde {\mathrm{d}s}_6^2}^2\,.
	\ee
	Then condition 1 requires that $\alpha F_3 +X_3$ scales as $\e^{4A}$. $F_3$ indeed has two legs on the NS5 when it threads the $A$-cycle, but we assume it cannot scale to zero near the NS5 brane for exactly the same reasons mentioned around equation (\ref*{chargedensity1}). 
	
	Hence either $\alpha$ and $X_3$ scale as $\e^{4A}$, or one does and the other one vanishes still more rapidly. We now argue that neither of these two possibilities can be excluded since sending $\alpha$ and $X_3$ to zero can still be consistent with a positive $1/\tilde v_4\oint_{\partial {\cal M}} \mathcal{B}$ integral. 
	
	Consider (\ref{9form2}) and its integral over the 9-surface in the IR along the NS5 horizon. Again one finds that the last term does not contribute by comparing with the known flat space solution. Interestingly the first two terms can be integrated explicitly, such that we find the equality
	\be\label{MADM2}
	M_{ADM} = \text{Vol}_4 \Bigl(\alpha_H Q_{3} + b_H Q_5 \text{Vol}_2\Bigl)\,,
	\ee
	where $Q_3$ is the \emph{monopole} anti-D3 charge and $Q_5$ the \emph{dipole} NS5 charge carried by the spherical NS5 brane defined as
	\begin{equation}
	 Q_3 = \int_{S^5} F_5 \,, \qquad Q_5 = \int_{S^3} H_3\,,
	\end{equation}
	for an $S^5$ surrounding the D3-branes and an $S^3$ linking the NS5-brane, as in Figure \ref{fig:Torus}.
	
	The values of the gauge potentials near the horizon are denoted
	\be
	C_4 = \alpha_H \tilde{\star}_4 1\,,\qquad B_6 = b_H \tilde{\star}_4 1\w \epsilon_2\,, 
	\ee
	with $\epsilon_2$ the volume-form of the 2-cycle wrapped by the NS5 brane, whose integrated value is denoted $\text{Vol}_2$. The above value formula for the ADM mass computed near the horizon coincides exactly with what would have been found using the formalism of \cite{Gautason:2013zw} (and \cite{Blaback:2014tfa}) that relates the ADM mass to the on-shell brane action $S_{DBI}  + S_{WZ}$. Indeed, the above expression is the on-shell Wess-Zumino term for a spherical NS5 brane carrying anti-D3 charge. The DBI probe action always vanishes on-shell since the warp-factor vanishes near the horizon. This is a nice consistency check of our results. It is now clear that we can take $\alpha_H$ to vanish as long as $b_H$ remains finite to account for the ADM mass. At the same time this is consistent with $X_3$ scaling down as $\e^{4A}$ such that the expression for the flux divergence (\ref{Hdensity3}) is consistent with a spherical NS5 brane. The way $X_3$ scales down is not affected by the finite value of $b_H$ (needed to account for the ADM mass) since $X_3$ relates to the derivatives of $b_H$ but not $b_H$ itself (\ref{dB6}). 
	
	We can glean interesting quantitative information from our analysis concerning the NS5 position as well. Consider the 2-cycle wrapped by the NS5. Its volume scales as $g_sM$
	\be
	\text{Vol}_2 = \int\epsilon_2 = g_sM R^2  \,, 
	\ee
	where $R$ is the radius of the $S^2$ on the A-cycle wrapped by the NS5 brane, with the overall $g_sM$ scaling taken out and we ignored numerical prefactors\footnote{This is a consequence of the fact that the $S^3$ tip of the KS cone has a radius set by $\sqrt{g_sM}$.}. Since the ADM mass is proportional to the number of anti-branes $p$ \cite{Dymarsky:2011pm,Junghans:2014xfa}
	\be
	M_{ADM} =  2\text{Vol}_4 \e^{4A_{\text{tip}}} p\,, 
	\ee
	we find from combining the last two equations with (\ref{MADM2}) that
	\be
	R^2\sim \frac{p}{M}\,.
	\ee
	This differs from the probe level, where it was found that for small $p/M$ \cite{Kachru:2002gs}
    \be
	R \sim \frac{p}{M}\,.
	\ee
    In the discussion we comment on the interpretation of this mismatch.

\begin{figure}[ht!]
	\begin{center}
	 \includegraphics[width=.4\textwidth]{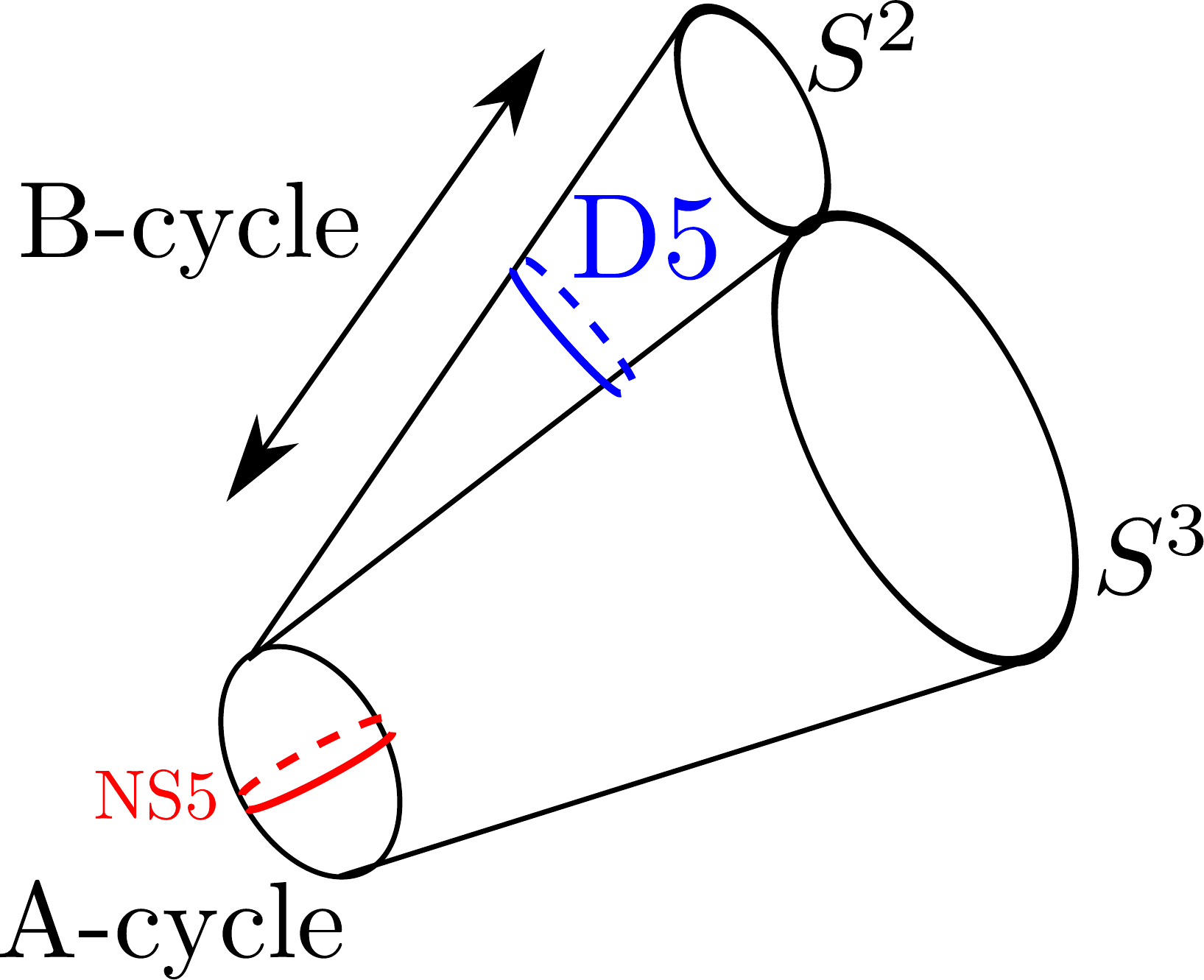}
	 \caption{Pictorial representation of the Klebanov-Strassler geometry. There are a priori two polarization channels for anti-D3 branes: NS5 polarization on an $S^2$ inside the A-cycle, or D5 polarization on the $S^2$ of the B-cycle.}
	 \label{fig:ABcycle}
	 \end{center}	
	\end{figure}

	Finally a word on the assumption of having a spherical NS5 brane as an IR configuration. This is clearly motivated by the probe analysis of \cite{Kachru:2002gs}. Nevertheless, it could be that the NS5 polarisation channel comes with partner five-brane polarisation channels when backreaction is taken into account, similar to the Polchinski-Strassler (PS) background \cite{Polchinski:2000uf}. Therefore, if the supergravity solution locally is a non-supersymmetric version of PS, there could be a rather involved web of $(p,q)$ 5-branes spanning different directions. The natural direction for D5 polarisation in the KS throat is the contractible B-cycle as drawn schematically in fig. \ref{fig:ABcycle}.
	If the D5 channel is also present, our computations are not valid for at least two reasons. First, having two intersecting branes around complicates the metric scaling such that our above arguments are invalidated. Second, we have relied on a trivial Bianchi identity for $F_3$, which is altered in the presence of D5 charges.
	It has been argued in the regime when $p/M \gg 1$ that D5 polarisation does not occur \cite{Bena:2012vz, Bena:2014jaa}. This seems to be a peculiar property of anti-branes. For the purpose of this paper, the absence of a D5-polarization channel also for small values of $p/M$ would imply a big simplification: we would then simply discard the analysis of boundary conditions where a spherical D5 polarization accompanies the NS5 brane.
	
	However the full story seems more intricate. As suggested in \cite{Bena:2014jaa} in the regime $p/M \gg 1$, the preferred channel might well be an oblique phase, a combination of D5 and NS5 polarisation along resp.\ B- and A-cycles. We leave a full treatment of possible oblique phases and more general asymptotics that allow to go beyond the limit $p/M\ll 1$ to future work.
	
	\section{Revisiting \texorpdfstring{$T>0$}{}  no-go claims}
	
	We go further in our investigation of the near brane behavior of anti-branes by heating up the background. If the metastable state persists in the supergravity regime, one expects to cover all the involved 5-brane polarisation processes behind a smooth horizon \cite{Freedman:2000xb}. For that reason we ignore the issue of not having a well defined $B_2$ field along the IR surface. 
	
	Once we introduce temperature in the form of a non-zero blackening factor, we find
	\begin{equation}
	e^{-\phi}|H_3|^2\sim e^{-2f}|\alpha F_3 + X_3|^2\,,
	\end{equation}
	where we  assume that $\e^{2A}$ and $\phi$ are finite at the horizon. It was shown in \cite{Blaback:2014tfa} that also at finite $T$ one can argue that $\alpha$ and $X_3$ cannot simultaneously  vanish at the horizon: this is necessary in order to have a non-zero finite boundary integral  $\oint \mathcal B$ at the horizon. Hence we seem to have a singular $H_3$ density since $e^{-2f}$ becomes infinite at the horizon.

	There is however one assumption that went into the no-go result of \cite{Blaback:2014tfa} that should perhaps be relaxed. Five assumptions were made explicitly. Let us take $r$ to be the coordinate perpendicular to the horizon:
	\begin{enumerate}
	  \item The temperature shows up under the form  of $g_{tt} \sim e^{2f}, g_{rr} \sim  e^{-2f}$, with $e^{2f} \sim r-r_h$.
	 \item $\e^{2A}, e^{2f},  \phi$ depend only on $r$ near the horizon  $r_h$.
	 \item The ansatz \eqref{H3ansatz} for metric and form  fields .
	 \item The component of $F_3$ with  all  legs along the horizon is non-zero.
	 \item  The relevant part  of the boundary term,
	\begin{equation}\label{boundaryint}
	\oint_{\partial\mathcal{M}_{horizon}}\tilde{\star}_4 1\wedge[\alpha F_5+B_2\wedge X_3]\,,
	\end{equation}
	is finite and non-zero.
	\end{enumerate}
	These assumptions are rather solid, following  either from black brane solutions (points 1 and 2), the expectation of the backreaction  in the KS background (points 3,4) or the study of the boundary term (5).  Condition 4 however was used in a stronger form: not only $F_3$, but $\alpha F_3+X_3$ was assumed to always have a component \textit{along} the horizon. This was partially inspired by former smeared branes setups in which this was necessary for preserving the symmetries \cite{Bena:2012ek}. We think that this requirement  might be too strong, and indeed some comments about this were already present in \cite{Blaback:2014tfa}. Our aim here is to further clarify this issue.
	
	First, one of the distinguishing features is that $F_3$ has flux along the topological A-cycle. This leads to $\star_6 H_3 = e^{\phi-4A-f} (\alpha F_3 + X_3)$ also having flux through this cycle. This $H_3$ flux  is the cause of the Myers effect that polarizes the probe branes into NS5-branes and we expect it to  be present in the backreaction as well.
	In general, the radial coordinate along the A-cycle will not be identified with the radial component  $r$ orthogonal to the horizon. Nevertheless, these coordinates will have a non-trivial relation, leading to the near-horizon behaviour:
	\be
	\alpha F_3 + X_3 = \mathrm{d}r \wedge \omega_2 + \ldots\,, 
	\ee
	for some two-form $\omega_2$. In  case the terms $\ldots$ vanish we find that the $H_3$ density remains finite since $g_{rr} \sim e^{-2f}$:
	\begin{equation}
	e^{-\phi}|H_3|^2\sim |\alpha F_3 + X_3|_{\widetilde{\mathrm{d}s}_6}^2\,,
	\end{equation}
	where now the contraction does not include $e^{2A}$ factors nor blackening factors $e^{2f}$. 
	Intriguingly, this is exactly the mechanism in \cite{Hartnett:2015oda} that provides a finite solution. However, it is unclear whether this extends to a viable  finite $T$ version of anti-D3 branes in KS backgrounds. In \cite{Hartnett:2015oda}, there was no topological compact A-cycle. Rather the simplification was made that the six-dimensional transverse space was $\mathbb{R}^6$, the role of the A-cycle being played by an $\mathbb{R}^3$ factor. Also, the work of  \cite{Hartnett:2015oda} included the first order backreaction of the fluxes on the black D3-brane geometry, but did not include the backreaction of the metric nor second order effects that can turn up in the flux density. We believe that this issue  still has to be settled.

\section{Conclusions and outlook}

In this paper we emphasized the importance of exploring  the boundary term that evaluates to the generalized ADM mass to shed new light on the back-reaction of anti-branes in flux throats. This method is very powerful, as it gives strong constraints on the back-reacted supergravity solutions without having to construct them explicitly. Our main result is the application of this method to the polarized NS5 state of supersymmetry-breaking anti-D3 branes in warped throats. Our method allows for the first time to study the polarized NS5 state in the regime where supergravity and probe limits are both expected to be applicable ($g_s p  \gg 1$ and  $p/M \ll 1$,  with $p$ the anti-brane charge and $M$ the 3-form flux through the A-cycle of the KS throat).

\subsection{Summary of results}
Our main result is the observation that the 3-form flux divergence, typical to anti-brane solutions, can potentially be made physical by polarising the anti-D3 branes into spherical NS5 branes. This polarisation process is expected from probe computations \cite{Kachru:2002gs} and the presence of local NS5 sources offers a natural explanation for the presence of singular 3-form fluxes. We found that the conditions for gluing the IR geometry to the UV geometry do not forbid such a 3-form singularity. Furthermore the computation suggests that there is a unique radius $R$ for the spherical NS5 brane and that the radius scales as $R\sim \sqrt{p/M}$. The probe analysis however suggests that $R \sim p/M$. So for small values of $p/M$, which is the regime in which one can expect meta-stable SUSY-breaking, the radius of the NS5 brane is much bigger than the radius predicted by the probe computation. This is a clear indication that, if the full back-reacted supergravity solution exists, the flux clumping process described in \cite{Blaback:2012nf, Danielsson:2014yga} indeed significantly pushes the NS5 brane towards the equator. However, in contrast with \cite{Blaback:2012nf, Danielsson:2014yga} we cannot conclude that it actually goes over the equator nor that the system becomes locally unstable.

Our result rests on the absence of D5 polarisation in the B-cycle of the KS throat, which was argued based on the computations carried out in \cite{Bena:2012vz, Bena:2014jaa}. This turns out to be the simplification needed to apply the techniques of \cite{Gautason:2013zw, Blaback:2014tfa} to compute the 3-form flux density near the source without the need for the full supergravity solution. Our proof does not depend on any details of the background (aside the absence of D5 polarisation): the only requirement is that supersymmetry is broken such that the (generalised) ADM mass is positive. But as mentioned in \cite{Bena:2014jaa} there can be oblique D5/NS5 polarisation channels and we consider it an interesting challenge to extend our result to that case.
	
As an aside, we reinvestigated the assumptions that went into the no-go theorem for the existence of smooth finite temperature anti-brane solutions \cite{Blaback:2014tfa}. In view of the possible NS5 polarisation channel we argued that a specific assumption about the directions of the 3-form flux near the horizon could be relaxed. If the 3-form $\star_6 H_3$  near the horizon is of the form $\mathrm{d} r \wedge \omega_2$, with $\omega_2$ a two-form and $r$ the local coordinate transverse to the horizon, the flux density at the horizon will be smooth. This condition cannot be satisfied for smeared anti-D3 solutions \cite{Bena:2012ek} or localised anti-D6 branes \cite{Bena:2013hr}. Hence if smooth finite $T$ solutions exist, then their construction will necessarily involve the boundary condition $\star_6 H =\mathrm{d} r\wedge \omega_2$ at the horizon, which can be natural for NS5 branes.

We used that at large enough $T$ we expect that the details of the NS5 polarization are hidden behind the horizon. One can still try to heat up the NS5-polarization itself to see the effect of small  temperature. We leave this for future work.


\subsection{Numerical analysis}

We hope that our analysis can be the starting point for a numerical investigation of fully backreacted NS5 solutions. Our results could be useful for choosing boundary conditions near the horizon in order to find a solution. We have shown that boundary conditions exist, consistent with the ADM mass, that evade unphysical singularities. Having certainty that anti-brane supersymmetry breaking is meta-stable as indicated by probe computations \cite{Kachru:2002gs} requires that a well-behaved supergravity solution can be found. At least this statement is true for large flux numbers and charges such that all typical length scales are within the classical gravity level. For small charges, arguments beyond the probe approximation have been suggested in \cite{Michel:2014lva, Polchinski:2015bea}.

If a numerical study suggests that our IR boundary conditions cannot be chosen, then an unphysical 3-form singularity remains that can only be interpreted as a fatal attraction of the D3 charges dissolved in flux towards the anti-D3 brane. In that case we speculate that this could be an explanation for the tachyon found in the analysis of Bena et.~al.~\cite{Bena:2014jaa}. The tachyon corresponds to a force on anti-D3 branes that has a non-zero projection towards the top of the A-cycle. The same should hence apply to spherical NS5 branes carrying anti-D3 charge \cite{Bena:2015kia}. If the picture of  \cite{Blaback:2012nf, Danielsson:2014yga, Gautason:2015ola} is correct, then one expects exactly a tachyonic mode that pushes spherical NS5 branes towards the North Pole of the $S^3$ A-cycle, consistent with \cite{Bena:2015kia}.\\

\subsection{Connection to Smarr relations and black hole physics} \label{ssec:Smarr}
In  \cite{Blaback:2014tfa}, the boundary term \eqref{9form} has only been related to the ADM mass for KS-like throats with added anti-brane charge of \cite{Dymarsky:2011pm} based on the perturbation of \cite{Bena:2009xk}. However, we believe that it can be applied much more widely. The expression for the mass  in terms of a boundary term that we used in this paper is a special case of a Smarr relation. For instance in five-dimensional supergravity, the generalized ADM mass for asymptotically flat or AdS solutions with horizons (black holes, black rings) can be derived from a boundary term that evaluates to (see \cite{Emparan:2006mm,Cvetic:2010jb} and references therein):
\begin{equation}
 M_{ADM} = T S  + \Phi Q + \phi q +  V \Lambda\,.\label{eq:Smarr}
\end{equation}
The parameters in this expression are charges and dual potentials: entropy $S$ and temperature $T$, electric monopole charge $Q$ and electrostatic potential $\Phi$, magnetic dipole charge $q$ and its magnetic potential  $\phi$, and (minus) the effective inside the horizon and the cosmological constant $\Lambda$ . The appearance of the dipole charge might be surprising: it is not a global conserved charge, but can only be measured locally for instance for dipole black rings. Still, it contributes to the Smarr relation because it is impossible to define the dipole potential $\phi$ using only a single patch, as  explained in great detail in  \cite{Copsey:2005se}. Recently the Smarr relation was shown to allow also for a bulk contribution (not a boundary term), that is non-zero only for non-trivial topology, which was previously overlooked. This topological contribution makes the construction of stationary spacetimes possible even in absence of horizons and underpins the black hole microstate geometry programme \cite{Gibbons:2013tqa}.

The form of the Smarr relation \eqref{eq:Smarr} is not restricted to asymptotically flat or AdS spaces, but to more general  asymptotics. The prescription for the conserved energy (mass) in generic spacetimes with a timelike Killing vector has been discussed by Hawking and Horowitz in a Hamiltonian formalism \cite{Hawking:1995fd}. Its application to flat space or AdS reduces to the known form of the Smarr formula in terms of Komar integrals, while for the anti-branes in KS it should become our boundary term \eqref{9form}.
We sketch the analogy between Smarr relations for black holes and warped throats in \cite{proceedings} and we will come back to this in great detail in a forthcoming publication \cite{futurework}.  

%
%
%
%
%


\section*{Acknowledgements}

We are happy to acknowledge useful discussions with J.\ Blaback, I.\ Bena, U.H.\ Danielsson, F.\ Gautason, G.\ Hartnett, D.\ Junghans, A.\ Puhm, B.\ Truijen D.\ Turton and especially J.\ Polchinski for pointing out a calculational error in the first version of this paper and the anonymous  JHEP referee for very  constructive feedback. TVR is supported by the National Science Foundation of Belgium (FWO) grant G.0.E52.14N Odysseus and Pegasus. We also acknowledge support from the European Science Foundation Holograv Network. The research of BV is supported by the European Commission through the Marie Curie Intra-European fellowship 328652--QM--sing. DCM would like to acknowledge the Becas Chile scholarship programme of the Chilean government.

\appendix{
\section{Appendix}

This appendix has three sections with more technical details. We rederive the boundary term that integrates to the ADM mass in \ref{app:boundary}, give a quick review of $p$-brane solutions in flat space  in secction \ref{app:sugra} and in section \ref{app:finiteT} we conclude with the details for extending our finite temperature results from anti-D3 to anti-D$p$ branes for any $p\leq 6$.

\subsection{Boundary term}\label{app:boundary}

Here we rederive the boundary term from the Einstein equations and other equations of motion. First we choose a shorthand notation $X_7$ in the ansatz that solves the $B_2$ equation of motion \eqref{B2EOM}:
\begin{equation}
 e^{-\phi} \star_{10} H_3 = -C_4  \wedge F_3 + X_7\,,\qquad {\rm d} X_7 = 0\,.
\end{equation}
In the main body of the text we used Poincar\'e invariance in four dimensions to write $X_7 = \tilde \star_4 1 \wedge X_3$  with $X_3$  closed.

With the ansatz we describe in section \ref{gluing}, the Einstein equations can be massaged to \cite{Blaback:2014tfa}
\begin{equation}\label{Ricci4}
R_{4}=-\nabla^2\phi-e^{-\phi} |H_3|^2-|F_{5}|^2-\mu_3 \delta(\Sigma).
\end{equation}
Here, $R_4$ is the trace of the Ricci tensor along the four macroscopic dimensions and $\Sigma$ denotes the brane world-volume. The ansatz \eqref{H3ansatz} together with the equations of motion for the form fields also imply that
\begin{align}
\star_{10}e^{-\phi}|H_3|^2&=-C_{4}\wedge F_{3}\wedge H_3+X_7\wedge H_3\,,\\
\label{F5}
\star_{10}|F_{5}|^2 &=\mathrm{d} (C_4\wedge F_5)+C_4\wedge F_3\wedge H_{3}- \mu_3\delta(\Sigma)C_4\wedge \star_6 1\,,\\
C_4& =\star_4 ( e^{-4A}\alpha)
\end{align}
From these equations, it follows that
\begin{equation}
\star_{10} R_{4}=\mathrm{d}\star_{10}\mathrm{d}\phi-\mathrm{d} (C_4\wedge F_5)-X_7\wedge H_3 +\star_{10}(\alpha e^{-4A}-1)\mu_3\delta(\Sigma).
\end{equation}
In order to get a still more suggestive form, we use the following relation between the Ricci scalars of the metrics with and without warp factors: 
\begin{equation}
R_{4}=e^{-2A}\tilde{R}_{4}+\star_{10}\text{d}\star_{10}\text{d}\left(4A\right).
\end{equation}
Then with \eqref{B6def} and $\tilde R_4 = 0$ (Minkowski space), we are left with
\begin{equation}\label{relation}
\star_{10}(1-\alpha e^{-4A})\mu_3\delta(\Sigma)=\mathrm{d}\star_{10}\mathrm{d}(\phi-4A)-\mathrm{d} (C_4\wedge F_5)-X_7\wedge H_3.
\end{equation} 
The first two terms on the right-hand side are total derivatives, but not the last one. However, we can remark that since $X_7$ is closed we can write it as
\begin{equation}
 X_7 = dB_6 + X_7^{\rm harm}\,,
\end{equation}
where $B_6$ is globally well-defined and $X_7^{\rm harm}$ is harmonic. From now on we assume that $X_7^{\rm harm} =0$. We will come back to the contribution of such harmonic terms to the ADM mass and the comparison to black hole microstate geometries and fuzzballs in future work. Then we can write
\begin{equation}
X_7\wedge H_3=\mathrm{d}(B_6\wedge H_3)-B_6\wedge \mathrm{d} H_3.
\end{equation}

Now, we integrate both sides of \eqref{relation} along a region of spacetime $\mathcal{M}$ not containing the source. The region $\mathcal{M}$ we have in mind has two boundaries: one  IR boundary surrounding the branes at a small distance, which we will let go to zero; and another one far in the UV of the KS throat. In the region 
${\cal M}$, there is no NS5 charge such that $\mathrm{d} H_3 = 0$ and we get
\begin{equation}\label{Boundaries}
\int_{\partial\mathcal{M}_{UV}}\mathcal{B}=\int_{\partial\mathcal{M}_{IR}}\mathcal{B}\,.
\end{equation}
where
\begin{equation}
\mathcal{B}=-C_4\wedge F_5-B_6\wedge H_3+\star_{10}\mathrm{d}(\phi-4A)\,.
\end{equation}

In \cite{Blaback:2014tfa} it was shown that in the UV the integral
\begin{equation}\label{adm}
\frac{1}{\tilde v_4}\int_{\partial\mathcal{M}_{UV}}\left(-C_4\wedge F_5+\mathrm{d} B_6\wedge B_2+\star_{10}\mathrm{d}(\phi-4A)\right)=M_{ADM}
\end{equation}
is equal to the ADM mass $M_{ADM}$, where $\tilde v_4$ is a volume factor accounting for the integration along the Minkowski directions. This ADM mass is non-vanishing whenever supersymmetry is broken by (anti-)D3 branes, as it is in the KPV set-up for metastable states. In the UV, $B_2$ can be integrated over $\partial {\cal M}_{\rm UV}$. By partial integration and taking into account the fact that we integrate along a boundary, we see that
\begin{equation}
\frac{1}{\tilde v_4}\int_{\partial\mathcal{M}_{UV}}\mathcal{B}= M_{ADM}\neq 0.
\end{equation}
When we combine this with \eqref{Boundaries} and realize that the dilaton and the warp factor do not contribute at the IR, we obtain the remarkable result
\begin{equation}\label{IRintegral}
\frac{1}{\tilde v_4}\int_{\partial\mathcal{M}_{IR}}\left(C_4\wedge F_5+B_6\wedge H_3\right)=M_{ADM}\neq 0.
\end{equation}
This is what we need to argue for the singularities in section \ref{divergences}.

\subsection{Supergravity brane solutions}\label{app:sugra}
Near the branes sources (D3 or NS5), we approximate the geometry by a $p$ brane metric:
\be
\mathrm{d} s^2 =  e^{2A} (-e^{-2f} \mathrm{d}t^2 + \mathrm{d}\vec x^2_{p}) + e^{2\frac{p+1}{p-7}A}(\mathrm{d}r^2 + r^2 \mathrm{d}\Omega^2_{8-p})\,,\label{eq:pbranemetric}
\ee
where the dilaton and sourced field strength for a D$p$ brane are
\begin{alignat}{2}
e^{\phi}&= e^{4\frac{3-p}{p-7}A}\,,\qquad &F_{8-p} = e^{2f}\coth \beta  Q_p \mathrm{d}\Omega_{8-p}
\end{alignat}
and for an NS5/F1 are
\begin{alignat}{2}
e^{\phi}&= e^{-4\frac{3-p}{p-7}A}\,,\qquad &H_{8-p} = e^{2f}\coth \beta  Q_p  \mathrm{d}\Omega_{8-p} \,.
\end{alignat}
Also,
\be
e^{\frac {16} {p-7}A} = 1 + \sinh^2 \beta r_0^{7-p}/r^{7-p}\,,\quad e^{2f}  = 1  - \frac{r_0^{7-p}}{r^{7-p}}\,.
\ee
The $T \to 0$ limit is $\beta \to \infty,r_0\to 0$ while keeping $Q_p \equiv \sinh^2 \beta r_0^{7-p}$ fixed.

Finally we note that at $T=0$, the flux density near $p$-branes scales with the warp factor $A$ as:
\begin{align}
NS5/F1:  \quad  & e^{(3-p)\phi/2}|H_{8-p}|^2 \sim  e^{-2\frac{(p-3)^2}{(7-p)^2} A}\,,\label{eq:singular_p_fluxes}\\
Dp:\quad &e^{-(3-p)\phi/2}|F_{8-p}|^2 \sim  e^{-2\frac{(p-3)^2}{(7-p)^2} A} \,.
\end{align}

\subsection{Anti-\texorpdfstring{D$p$}{} branes at finite \texorpdfstring{$T$}{}}\label{app:finiteT}
The extension of the finite temperature results to anti-D$p$ branes inserted in throat geometries that carry D$p$ brane charges dissolved in fluxes is immediate \cite{Blaback:2014tfa}. The ansatz generalises to
\begin{align}
&\mathrm{d} s^2_{10} = \e^{2A} g_{\mu\nu} \mathrm{d} x^\mu \mathrm{d} x^\nu + \mathrm{d} s^2_{9-p}\,, \\
& C_{p+1}= \tilde \star_{p+1} \alpha ,\nonumber \\
& H_3 = \e^{\phi-(p+1)A} \star_{9-p}\Bigl(\alpha F_{6-p} +  X_{6-p}\Bigr)\nonumber\,.
\end{align}
Such backgrounds exist up to $p=6$, which describes an anti-D6 brane in a background with Romans mass and $H_3$ flux carrying D6 charges. As with anti-D3 branes, there is a conserved current that entails a non-trivial gluing condition between the IR and the UV. Near the horizon in the IR the conserved charge is given by
\be
\oint_{IR}\frac{2}{1-p}\tilde{\star_{p+1}}1\w[ F_{8-p} + B_2\w X_3] + \frac{4}{p+1}\star_{10}\mathrm{d} f\,,
\ee
and as before this integral has to be positive and finite. 

In complete analogy with anti-D3 branes we can investigate necessary gluing conditions to obtain the same conclusions, except for anti-D6 branes where we find a different result. When $p=6$, the three-form $H_3$ is always proportional to the volume form in the three-dimensional transverse space (there is no A-cycle) and there is no $X_0$ term: $H_3 = e^{\phi-7A} \star_3 F_0$.  For finite $T$ solutions we furthermore cannot evade the no-go result as sketched earlier, since  $\star_3 H_3 \sim F_0$ is a zero-form and hence cannot have a direction along $\mathrm{d} r$. 
Therefore we still find a singular horizon for anti-D6 branes in flux background and their T-dual equivalents such as anti-D3 branes smeared on the tip of the Klebanov-Strassler throat.

}

%
%

\providecommand{\href}[2]{#2}\begingroup\raggedright\endgroup

\end{document}